\def\bey{\begin{eqnarray}}
\def\eey{\end{eqnarray}}
\def\be{\begin{equation}}
\def\ee{\end{equation}}
\def\ba{\begin{array}}
\def\ea{\end{array}}

\def\gm{\gamma}

\def\ld{\lambda}

\def\sg{\sigma}

\def\r{\rho}

\def\pp{\partial}

\def\ra{\rightarrow}
\def\nnb{\nonumber}
\documentstyle[11pt,epsfig]{article}
\title{  Is the temperature as a Lorentz scalar?}
\author{ W.Z.Jiang
\footnote{  Email address: jiangwz02@hotmail.com}\\
\hspace{-0cm}$^1$ Shanghai Institute of Nuclear Research\\
  Chinese Academy of Sciences, Shanghai 201800, China\\
$^2$ Center of Theoretical Nuclear Physics,\\
 National  Laboratory of Heavy Ion Accelerator,
Lanzhou 730000, China}
\date{}
\begin{document}
\maketitle \baselineskip 20.6pt
\begin{abstract}
\bigskip
\baselineskip 18pt  The thermodynamic interaction of a free
Fermion system is alternatively described by the coupling of
particles with the vacuum of a scalar pseudo boson which features
a self-interaction. The proper temperature is shown as a Lorentz
scalar in a Lorentz invariant framework. At thermodynamic
equilibrium, the Lorentz transformation of temperature is
investigated based on the conceptions of the relativistic quantum
field theory.



\end{abstract}

 \newpage

In thermodynamics, it has been long  controversial  whether the
temperature which represents the averaged kinetic energy is a
scalar or a component of the
vector\cite{haar,balescu,kraizan,lifl,strel,aldro,costa,fenech}.
Particles in a moving system obey laws of the relativistic
statistical physics in which the Lorentz factor is introduced. Two
years after Einstein's fundamental paper in 1905 on the
electrodynamics of moving bodies, Einstein\cite{einstein} and
Planck et.al.\cite{planck} deduced the transformation of the
temperature of a thermal body observed in  an inertial frame $S$
moving with constant velocity $v$ with respect to the center of
mass frame (rest frame) $S_0$
\begin{equation}
 T=T_0/\gm
 \label{eqt1}
\end{equation}
where $T_0$ is the proper temperature at the center of mass (rest)
frame, T is the apparent temperature of the moving body measured
in the laboratory frame, and $\gm=1/\sqrt{1-v^2}$ is the Lorentz
factor. It implies that a moving body appears to be cool. This
appeared to be the standard result\cite{pauli,tolman}. The
controversy arose as a different transformation of temperature had
been derived by Eidington\cite{eidington}, Ott\cite{ott} and
M$\phi$ller\cite{muller} based on the respective footings. In
their version, the transformation is as follows
\begin{equation}
 T=T_0\gm
 \label{eqt2}
\end{equation}
Landsberg\cite{lands} illustrated the temperature as a Lorentz
scalar and consequently redefined the first law of thermodynamics
for moving bodies. Kampen\cite{kampen} also illustrated that it is
better to take the temperature as a scalar.

The measurement of the cosmic microwave background in specific
frames has been drawn great attentions. The Unruh-DeWitt
detector\cite{unruh} that moves in a background thermal bath was
proposed to measure the relation of temperature
transformation\cite{costa}. However, final results would be
dependent on the temperature-defining and -measuring
procedures\cite{aldro,costa}. Based on  the theoretical
investigation of the black body radiation using the Unruh-DeWitt
detector, it was concluded\cite{land0} that the proper temperature
alone is left as the only temperature of universal significance.

One may doubt on whether the proper temperature at the rest frame
is a Lorentz scalar, since the temperature definition based on the
relation of the heat and entropy may suggest that the temperature
be not a Lorentz scalar.  The answer may turn out to be
affirmative according to the Ott's derivation of temperature
transformation\cite{ott}. Actually,  the mass modified by the heat
added is still the proper mass at the rest frame, which is the
first Senario (I) in this work. Due to the indefinition of the
apparent temperature, it is necessary to investigate what
principles, definitions, or relations for the temperature and
related thermodynamic quantities are obeyed under the Lorentz
boost. In order to serve this purpose, we investigate the thermal
system of the free Fermions (except for the thermodynamic
interactions) which obey the Lorentz invariant Dirac equation.
Before going into details, we give the second Senario (II) of this
work: the thermodynamic equilibrium is Lorentz invariant, which
was once took by Aldrovandi\cite{aldro}.

The non-linearity due to the  bremsstrahlung processes in the
thermal environment is beyond the kinetics demonstrated in the
Dirac equation. Some dynamical degrees of freedom should be
introduced.  According to the notion of modern field theory, the
interaction is mediated by medium bosons. Based on this notion,
one may introduce appropriate medium bosons to describe
thermodynamic interactions for the Fermion gas at thermodynamic
equilibrium. Note that the emphasis is placed on the thermodynamic
equilibrium. The most favorable bosons are the scalar and vector
ones. The vector boson couples to the non-zero and conserved
thermodynamic current which does not exist in the system at
thermodynamic equilibrium. One may argue that the temporal
component of the vector boson may exist. However, such an
existence of the temporal component contradicts with the Senario
II, since the thermal current appears as performing the Lorentz
boost. Consequently, only the scalar boson mediates the
thermodynamic interaction. The coupling of the scalar boson with
the Fermion is equivalent to the heating which increases the
Fermion mass. Hence, the scalar boson has to be of the imaginary
mass, based on the knowledge of the mean-field treatment. The
boson with the imaginary mass is however unphysical.  In order to
obtain the real boson mass, the self-interaction has to be
introduced. This is quite similar to the property of the Higgs
field\cite{higgs}. Such a boson is called as the scalar pseudo
boson because it leads to the increase but not the decrease of the
Fermion mass like the conventional boson.

The free scalar pseudo boson field, understood as the
thermodynamic field, obeys the following Lagrangian
\begin{equation}
{\cal L}_\tau=\frac{1}{2}
(\pp_\mu\phi_\tau\pp^\mu\phi_\tau-\mu_\tau^2\phi_\tau^2)
-\frac{\ld}{4}\phi_\tau^4\label{lag1}
\end{equation}
where  $\mu_\tau$ and $\ld$ are the constants with $\mu_\tau^2<0$
and $\ld>0$. The Hamiltonian is thus given as
\begin{equation}
 {\cal H}_\tau= \frac{1}{2}
(\dot{\phi}^2_\tau+(\nabla\phi_\tau)^2+\mu_\tau^2\phi_\tau^2)
+\frac{\ld}{4}\phi_\tau^4 \end{equation}
 Yet, we need to obtain the real mass of $\phi_\tau$.
The constant vacuum field quantity is determined by minimizing the
Hamiltonian,
\begin{equation} \phi_\tau^0=\pm\sqrt{\frac{-\mu^2_\tau}{\ld}},
\hbox{ }{\cal H}_\tau^0=-\frac{\mu^4_\tau}{4\ld}
\end{equation}
It is obvious that the constant vacuum field is symmetry-breaking
under the reflection of internal field space.  The free
thermodynamic field has the negative minimum energy. Performing
the substitution \begin{equation}
\phi_\tau=\phi_\tau^0+\phi=\sqrt{\frac{-\mu^2_\tau}{\ld}}+\phi
\label{higg}
\end{equation}
into Eq.~\ref{lag1}, it has the following form
\begin{equation}
{\cal L}_\tau=
\frac{1}{2}(\pp_\mu\phi\pp^\mu\phi-m_\tau^2\phi^2)-\ld
\phi_\tau^0\phi^3 -\frac{\ld}{4}\phi^4 \label{lag2}
\end{equation}
where the real mass of $\phi$ is obtained to be
$m_\tau=\sqrt{-2\mu^2_\tau}$. As seen in Eq.\ref{higg}, the
minimum of the thermodynamic field (i.e. the vacuum) is actually
redefined and it is now symmetric under field reflection. However,
the symmetry of the Lagrangian ${\cal L}_\tau$ in Eq.\ref{lag2} is
broken due to the term $-\ld \phi_\tau^0\phi^3$.

The total Lagrangian which contains the Fermion, the scalar pseudo
boson and the Yukawa coupling between them is in the following
form
 \bey {\cal L} &=&\bar{\psi}(i\gm_\mu \pp^\mu -m-g_\tau \phi_\tau)
\psi\nnb\\
&&+\frac{1}{2}(\pp_\mu\phi\pp^\mu\phi-m_\tau^2\phi^2)-\ld
\phi_\tau^0\phi^3 -\frac{\ld}{4}\phi^4 \label{lag3}\eey where
$\psi$ is the Dirac spinor of Fermion with the mass $m$. Based on
the viewpoint that thermodynamic equilibrium comes from energy
exchanges by the arbitrarily multiple collisions and multiple
radiation, let's see whether such a scalar coupling is enough to
contain all possible processes. The interaction part of the above
Lagrangian is as follows:
\begin{equation}
{\cal L}_I= -\bar{\psi}g_s S_\tau\psi=-\bar{\psi}g_s
e^{\sg_\tau}\psi=- \bar{\psi}\sum_{n=0}^{\infty}g_s
\frac{\sg_\tau^n}{n!}\psi\label{lag4}
\end{equation}
where  $S_\tau=\phi_\tau{\cal T}$ and  $g_s=g_\tau/{\cal T}$ with
${\cal T}$ the constant parameter in unit of space-time. The
auxiliary quantum $\sg_\tau$ is dimensionless. Assuming the
expansion power $n$ may represent the multiplicity of the
interaction processes, the interaction potential $g_\tau
\phi_\tau$ can be understood as an averaged quantity over all
multiple processes with the same probability within a certain
characteristic spacing of space or time.

The scalar field $\phi$ satisfies following equation
\begin{equation}
(\pp^2+m_\tau^2)\phi=-g_\tau\bar{\psi}\psi-3\ld\phi_\tau^0\phi^2
-\ld\phi^3 \label{field1}
\end{equation}
In homogeneous thermal matter, it has
\begin{equation}
\phi=\frac{1}{m^2_\tau}(-g_\tau\bar{\psi}\psi-3\ld\phi_\tau^0\phi^2
-\ld\phi^3) \label{field2}
\end{equation}
In an alternative description for the thermodynamic interaction,
the real-mass boson does not always suggest the definite existence
of the boson. Actually, a scalar boson which serves the role to
like the Higgs boson is not found yet. It may be hence reasonable
to set the boson mass as $m_\tau\ra \infty$ in this study, which
will be justified below. This means that it has $\phi\ra 0$. The
thermal heat is hence added to the system just by the term
$g_\tau\bar{\psi}\phi_\tau^0\psi$ in Eq.\ref{lag3}. The coupling
with the isotropic field vacuum $\phi_\tau^0$ can nicely be used
to elaborate the homogeneous thermodynamic equilibrium of the
Fermion system. It is actually the broken vacuum of the scalar
field that plays the very role, and that is quite similar to a
precedent, the Higgs boson in the gauge theory.

The coupling potential $g_\tau\phi_\tau^0$ can be evaluated from
the Senario (I). According to the equipartition theorem, the
Fermion mass measured at the center of mass frame at the low
temperature limit is in the following
\begin{equation}
m^*=m+g_\tau\phi_\tau^0=m+\frac{3}{2}T_0 \label{eqt3}
\end{equation}
 The Dirac equation is then given as
\begin{equation}
[i\gm_\mu \pp^\mu -(m+3T_0/2)] \psi=0\label{dirac}
\end{equation}
which indicates that the proper temperature $T_0$ modifying the
proper mass of Fermions is a Lorentz scalar. The derivation of
Eq.\ref{eqt3} contains two implications. First, there is the
additivity of the thermal energy for the particle and
anti-particle, which is  consistent with that the scalar density
is obtained from the sum, other than the difference, of the
particle and anti-particle distributions.  Second, it is
reasonable to take $m_\tau\ra\infty$ above. Otherwise, part of the
thermal energy, which is from the nonlinear term of the  field
$\phi$ in Eq.\ref{lag3}, will not be carried by the Fermions,
contradicting with the relation given by Eq.\ref{eqt3} based on
the equipartition theorem.  Even if the nonlinear term gives an
infinitesimal contribution, it will lead to the inequality
$m^*\neq m+3/2T_0$ for finite $m_\tau$.

We recall that we have excluded the existence of the vector boson
by the field theory at thermodynamic equilibrium, which is
necessary to obtain a proper temperature as a Lorentz scalar other
than the component of a vector in a field language.  With the
above formalism, the thermodynamic interaction is alternatively
described by the dynamic scalar boson coupling for a thermal but
free Fermion system. It is worthwhile to note that the single
Fermion property in the thermal environment is treated in the
sense of the statistical average or normalization in the total
phase space involved.

Now it is convenient to discuss the transformation properties for
the apparent temperature based on the Lorentz structure of
Eqs.\ref{lag3} and \ref{dirac}. It is able to obtain various
apparent temperature without violating the relativity principle
and the structure of the Dirac equation. There is the following
relation for the total energy $E$ of one particle in a co-moving
system
\begin{equation}
E=\sqrt{p^2+{m^*}^2}=\gm m^*=\gm m+\frac{3}{2}T_0\gm
\end{equation}
where the apparent temperature is given as $T=T_0\gm $, which is
the result of Eq.\ref{eqt2}\cite{eidington,ott,muller}. On the
other hand,  we note that the density of thermal heat $\r_\tau=
g_\tau\bar \psi\phi_\tau^0 \psi$ is a Lorentz scalar, while the
total thermal heat $Q=\int d^3x\r_\tau$ is not a Lorentz scalar
due to the volume element $d^3x$. According to the definition
$T=\pp Q/\pp S$ with $S$ the entropy, it has the relation
$T=T_0/\gm$ which is the result of
Eq.\ref{eqt1}\cite{einstein,planck}. The distinct difference
arises from the different measurement. The former is by means of
the mass measurement, while the latter is through measuring the
volume. The different measurement actually corresponds to
different definitions or deduction of the apparent temperature. It
is important to make clear what definitions and their relations of
thermodynamic quantities are used for measurement and deduction.
If one performs a coincident mass and volume measurement for the
quantity $(m^*\cdot Q)$, the apparent temperature is then $T=T_0$.
More coincident measurements can be performed under various
combinations.

In summary,  the thermodynamic motion of a free Fermion gas is
illustrated in a dynamic framework where the thermodynamic
interaction is alternatively described by a scalar boson exchange
based on notions of the relativistic field theory and the senario
for thermodynamic equilibrium. The boson field has the
self-interaction which gives rise to its real mass. The
homogeneous coupling of the Fermion with the broken vacuum of the
scalar field modifies the Fermion mass. The coupling potental is
worked out according to two senarios, justifying the proper
temperature in the free Fermion gas as a Lorentz scalar. The
measurement dependence of the apparent temperature have been also
investigated based on the relativity principle and the Lorentz
structure of the formalism. The controversy for different apparent
temperature seems to come from the different definition and
measurement.

\section*{Acknowledgement}
The author  thanks Dr. X.B.Ai  and  Dr. F.Kleefeld for useful
communications. This work is partially  supported by CAS knowledge
Innovation Project No.KJCX2-N11  and the Major State Basic
Research Development Program Under Contract No. G200077400.


\begin{thebibliography}{50}
\bibitem{haar}D.ter Haar, H.Wergeland, Phys. Rept. 1(1971)31
\bibitem{balescu}R.Balescu, Physica 40(1968)309
\bibitem{kraizan}J.E.Kraizan, Phys. Lett. A 71(1979)174
\bibitem{lifl}F.L.Li, Prog. Phys.(in Chinese)8(1989)362
\bibitem{strel}V.N.Strel'tsov, JINR-D2-91-367, (1992) (KEK scanned
version)
\bibitem{aldro}R.Aldrovandi, J.Gariel, Phys. Lett. A 170(1992)5
\bibitem{costa}S.S.Costa, G.E.A.Matsas, Phys. Lett. A 209(1995)155
\bibitem{fenech}Ch.Fenech, J.P.Vigier, Phys. Lett. A 215(1996)247
\bibitem{einstein}A.Einstein, Jahrbuch d. Radioaktivitat und
Elektronik, 4(1907)411
\bibitem{planck}M.Planck, S.B.Preuss, Akak. Wiss (1907)542
\bibitem{pauli}W.Pauli, Theory of Relativity (Pergamon Press,
London,1958)
\bibitem{tolman}R.C.Tolman, Relativity, Thermodynamics and
Cosmology (Oxford University Press,1934)
\bibitem{eidington}A.S.Eidington, The Mathematical theory of
relativity, (Cambridge Press, 1957)
\bibitem{ott}H.Ott, Z. Phys. 70(1963)175
\bibitem{muller}G.M$\phi$ller, Relativistic thermodynamics,
(K$\phi$benhavn, 1967)
\bibitem{lands}P.T.Landsberg, Nature, 212 (1966)571; ibid,
214 (1967)903
\bibitem{kampen}N.G.Van Kampen, Phys. Rev. 173(1968)295
\bibitem{unruh}W.G.Unruch, Phys.Rev. D14(1976)870; B.S. DeWitt,
General relativity, ed. S.W.Hawking and W.Israel (Cambridge
University press, 1979)
\bibitem{land0}P.T.Landsberg, G.E.A. Matsas, Phys. Lett. A
223(1996)401
\bibitem{higgs}P.W.Higgs, Phys. Rev. 145(1966)1156
\end{thebibliography}
\end{document}